\documentclass{PoS}
\usepackage{slashed}
\usepackage{amsmath,amsthm, amssymb, latexsym}

\title{Contribution to the neutrino magnetic moment coming from 2HDM in
presence of magnetic fields.}

\ShortTitle{Contribution to MDM coming from 2HDM to neutrinos in a magnetic fields}

\author{ \speaker{Carlos G. Tarazona}$^{,a,b,1}$,Rodolfo A. Diaz $^{a,2}$, John Morales$^{a,3}$
and Andr\'es Castillo$^{a,4}$\\
\llap{$^{a}$}Universidad Nacional de Colombia, Sede Bogot\'a, Facultad de Ciencias,
Departamento de F\'{\i}sica. Ciudad Universitaria 111321, Bogot\'a, Colombia\\
\llap{$^{b}$}Departamento de Ciencias B\'asicas. Universidad Manuela Beltr\'an.
Bogot\'a, Colombia\\
E-mail: \email{$^{1}$caragomezt@unal.edu.co}, \email{$^{2}$radiazs@unal.edu.co},
\email{$^{3}$jmoralesa@unal.edu.co}, \email{$^{4}$afcastillor@unal.edu.co}}

\abstract{The confirmation of the neutrino mass by oscillation phenomena converts the study of the
magnetic dipole moment (MDM) of the neutrino, in vacuum and regions where existing external
magnetic fields, a topic of particular interest from the theoretical point of view. 
The MDM has an implicit relation with neutrino masses, and this is a possible benchmark 
from new physics in the solution of open questions in neutrino physics. Besides we know 
that this kind of phenomena has significant consequences on cosmology and astrophysics, e.g., 
under the influence of combined effects of neutrinos in the compact objects formation and 
evolution of primordials magnetic fields. \
We calculate and analyze such effects introducing charged Higgs bosons based on the parameter 
space of several 2HDMs with and without flavor conservation in neutral currents.}

\FullConference{38th International Conference on High Energy Physics\\
		3-10 August 2016\\
		Chicago, USA}

\begin{document}

\vspace{-0.8cm}

\section{The electromagnetic form factors (EFF's)\label{form factors}}

To examine the electromagnetic form factors is necessary
analyzing the interaction of the particle with the photon. The general
expression to the electromagnetic current in the vertex $\Lambda _{\mu
}\left( l,q\right) $ is \cite{Marek}:
\begin{equation}
\Lambda _{\mu }\left( l,q\right) =F_{Q}\left( q^{2}\right) \gamma _{\mu
}+F_{M}\left( q^{2}\right) i\sigma _{\mu \nu }q^{\nu }+F_{E}\left(
q^{2}\right) \sigma _{\mu \nu }q^{\nu }\gamma _{5}+F_{A}\left( q^{2}\right)
\left( q^{2}\gamma _{\mu }-q_{\mu }\slashed{q}\right) \gamma _{5},
\label{EFF}
\end{equation}
where $F_{Q}\left( q^{2}\right) $ corresponds to the factor of electric
charge, $F_{M}\left( q^{2}\right) $ is associated with the anomalous magnetic
moment, $F_{E}\left( q^{2}\right) $ is the electric dipole moment, and $%
F_{A}\left( q^{2}\right) $ is called anapolar moment. When we consider
neutrino masses, the MDM value of neutrino in the vacuum is $\mu _{\nu
_{\alpha }}=3.2\times 10^{-19}\mu _{B}\left( {m_{\nu _{\alpha }}}/{1 \text{ eV}}%
\right) $, where $\mu _{B}={e}/{2m_{e}}$ is the Bohr magneton.

\begin{figure}[tph]
\centering
\includegraphics[scale=0.9]{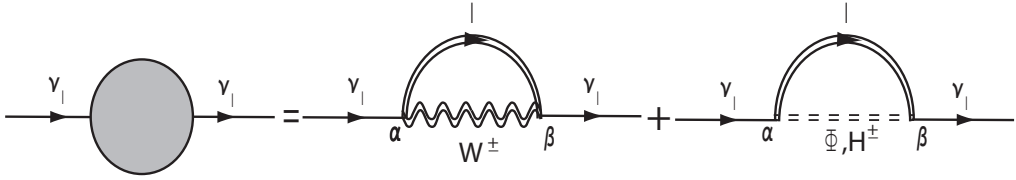} \vspace*{-0.4cm}
\caption{\textit{Feynman diagram representing the contribution to
self-energy due to a constant and uniform magnetic field, the double line
corresponds to lepton charged }$l$\emph{, the }$W^{\pm }$\emph{\ boson, the
charged Goldstone boson }$\Phi $\emph{, and the charged Higgs boson $H^{\pm }$\emph\, in an external magnetic field.}}
\label{self_ener}
\end{figure}

When we take into account an external magnetic field with massive neutrinos, the computation
of the electromagnetic form factors can make it by the self energy compation using the 
proper-time (Schwinger) formalism \cite{Erdas}. The Feynman diagrams contributing to the neutrino 
Self-Energy are depicted in figure \ref{self_ener}. The double lines represent the interaction with 
the magnetic field.

\vspace{-0.3cm}

The new structure to the gauge bosons $G_{B}\left( p\right) $, Goldstone
bosons $D\left( p\right) $, Higgs charged $D_{H}\left( p\right) $ and
fermions $S_{B}^{F}\left( p\right) $ propagators due to the presence of an
external magnetic field are described respectively by
\begin{eqnarray}
G_{B}\left( p\right)  &=&-\frac{ig_{\alpha \beta }}{p^{2}-m_{W}^{2}}-\frac{%
2\beta \varphi }{\left( p^{2}-m_{W}^{2}\right) ^{2}}+\mathcal{O}\left( \beta
^{2}\right) ,~\ D\left( p\right) =\frac{i}{p^{2}-m_{W}^{2}}+\mathcal{O}%
\left( \beta ^{2}\right) , \notag \\
D_{H}\left( p\right)  &=&\frac{i}{p^{2}-m_{H^{\pm }}^{2}}+\mathcal{O}\left(
\beta ^{2}\right) ,~\ S_{B}^{F}\left( p\right) =\frac{i\left( m-{\slashed{p}}%
\right) }{p^{2}-m^{2}}+\beta \frac{\left( m-{\slashed{p}}_{\parallel }\right) }{%
2\left( p^{2}-m^{2}\right) ^{2}}\left( \gamma \varphi \gamma \right) +%
\mathcal{O}\left( \beta ^{2}\right).  \label{propagadores}
\end{eqnarray}
where $\beta =eB$ and $\ \varphi ^{\alpha \eta }=F^{\alpha \eta }/B$  is the
dimensionless electromagnetic field tensor normalized to $B$, with the
Lorentz indices of tensors are contracted as $\gamma \varphi \gamma =\gamma
_{\alpha }\varphi ^{\alpha \beta }\gamma _{\beta }$ and dual tensor $\tilde{%
\varphi}^{\alpha \eta }=\frac{1}{2}\epsilon ^{\alpha \eta \zeta \vartheta
}\varphi _{\zeta \vartheta }$. The self-energy operator has the following Lorentz structure 
\cite{Erdas} 
\begin{equation}
\sum \left( p\right) =\left[ a_{L}\slashed{p}+b_{L}\slashed{p}_{\parallel
}+c_{L}\left( p\tilde{\varphi}\gamma \right) \right] P_{L}+\left[ a_{R}\slashed{p%
}+b_{R}\slashed{p}_{\parallel }+c_{R}\left( p\tilde{\varphi}\gamma \right) %
\right] P_{R}+m_{v}\left[ K_{1}+iK_{2}\left( \gamma \varphi \gamma \right) %
\right]. 
\label{MDM_self}
\end{equation}%
where $p^{\mu }=p_{\parallel }^{\mu }+p_{\perp }^{\mu }=\left(
p^{0},0,0,p^{3}\right) +\left( 0,p^{1},p^{2},0\right) $. The terms that represent the contibution to MDM have the structure \cite{Dobrynina}%
\begin{equation}
\mu _{\nu _{l}}^{B}=\frac{m_{\nu }}{2B}\left( c_{L}-c_{R}+4K_{2}\right) ,
\label{MDM_gen}
\end{equation}%
the result of this contribution calculated by the self-energy is
\begin{equation}
\mu _{\nu _{l}}^{B}=\mu _{\nu _{l}}\frac{1}{\left( 1-\lambda _{l}\right) ^{3}%
}\left( 1-\frac{7}{2}\lambda _{l}+3\lambda _{l}^{2}-\lambda _{l}^{2}\ln
\lambda _{l}-\frac{1}{2}\lambda _{l}^{3}\right), 
\label{MDM_result}
\end{equation}%
where $\lambda _{l}={m_{l}^{2}}/{m_{W}^{2}}$. 
The MDM in the vacuum and in presence of a magnetic field are described in table \ref{tab:EMNC}.

\vspace{-0.4cm}

\begin{table}[tbph]
\begin{center}
\begin{tabular}{|l|l|l|l|}
\hline
& $m_{\nu }\left[ eV\right] $ & $\mu _{\nu }\left[ \mu _{B}\right] $ & $\mu
_{\nu }^{B}\left[ \mu _{B}\right] $ \\ \hline
$\nu _{e}$ & $0.06089$ & $1.948\times 10^{-20}$ & $1.948\times 10^{-20}$ \\ 
\hline
$\nu _{\mu }$ & $0.06754$ & $2.161\times 10^{-20}$ & $2.161\times 10^{-20}$
\\ \hline
$\nu _{\tau }$ & $0.07147$ & $2.287\times 10^{-20}$ & $2.286\times 10^{-20}$
\\ \hline
\end{tabular}%
\end{center}
\vspace{-0.6cm}
\caption{\textit{MDM of the neutrinos in the vacuum and with magnetic fields. 
The values of the effective flavor masses are compatible with cosmological bounds and 
the masses for eigenstates in normal ordering.}}
\label{tab:EMNC}
\end{table}

\vspace{-0.6cm}

\section{Contributions to MDM comming from 2HDM with $\vec{B}$\label{2HDM}}
To include the contribution to MDM owing to 2HDM, we take into account the RS of 
fig. \ref{self_ener}. Using the propagators of Eqs. (\ref{propagadores}) and 
writing the respective vertices like $aP_{L}+bP_{R}$, we got
\begin{equation}
\sum\nolimits_{H^{\pm }}\left( p\right) =i\int \frac{d^{4}k}{\left( 2\pi
\right) ^{4}}\left( aP_{L}+bP_{R}\right) S_{B}^{F}\left( p-k\right) \left(
cP_{R}+dP_{L}\right) D_{B}\left( k\right).
\label{MDM_2hdm}
\end{equation}%
Factorizing the terms $c_{L},c_{R}$ and $K_{2}$ like in Eq. (\ref{MDM_gen}), 
the contribution to the MDM of neutrinos is 
\begin{equation}
\left( \mu _{\nu _{l}}^{B}\right) _{H^{\pm }}=\frac{\sqrt{2}}{3G_{f}}\mu
_{\nu _{l}}\int\limits_{0}^{1}dx\frac{2\left( x^{2}-3x+2\right) \left(
b^{2}+a^{2}\right) +\left( 1-x\right) \frac{m_{l}}{m_{\nu }}ab}{m_{\nu
}^{2}x^{2}+\left( m_{l}^{2}-m_{\nu }^{2}-m_{H^{\pm }}^{2}\right) x+m_{H^{\pm
}}^{2}},
\label{MDM_2hdm_result}
\end{equation}%
where the constants $a$ and $b$ are model dependent as showed in table \ref{tab:ab}. 
The results of Eq.(\ref{MDM_2hdm_result}) are displayed in figure \ref{type_I_II}.

\vspace{-0.5cm}

\begin{table}[tbph]
\begin{center}
\begin{equation*}
\begin{tabular}{|l|c|c|c|}
\hline
Vertex & Type I or Flipped 2HDMs & Type II or Lepton Specific 2HDMs & Type III 2HDM \\ \hline\hline
$a$ & $-2^{\frac{3}{4}}\sqrt{G_{F}}m_{\nu _{l}}\cot \beta U_{k,i}$ & $2^{%
\frac{3}{4}}\sqrt{G_{F}}m_{\nu _{l}}\cot \beta U_{k,i}$ & $-\lambda _{\nu
}\xi _{i,k}^{\nu }U_{k,i}$ \\ \hline
$b$ & $2^{\frac{3}{4}}\sqrt{G_{F}}m_{l}\cot \beta U_{i,k}$ & $2^{\frac{3}{4}}%
\sqrt{G_{F}}m_{l}\tan \beta U_{i,k}$ & $\lambda _{l}U_{i,k}\xi _{k,i}^{E}$
\\ \hline
\end{tabular}%
\end{equation*}%
\end{center}
\vspace{-0.6cm}
\caption{\textit{Couplings for $P_{L}$ and $P_{R}$ in the MDM for type I (Flipped), II 
(Lepton Specific), III-2HDMs.}}
\label{tab:ab}
\end{table}

\vspace{-0.5cm}

\section{Concluding remarks\label{sec:conclusions}}
\begin{figure}[tph]
\centering
\textbf{{\small Type I (Flipped) and II (Lepton-specific) \hspace{2.0cm}
Type III (Sher and Cheng Anzats)}}\newline
\includegraphics[width=0.48\columnwidth]{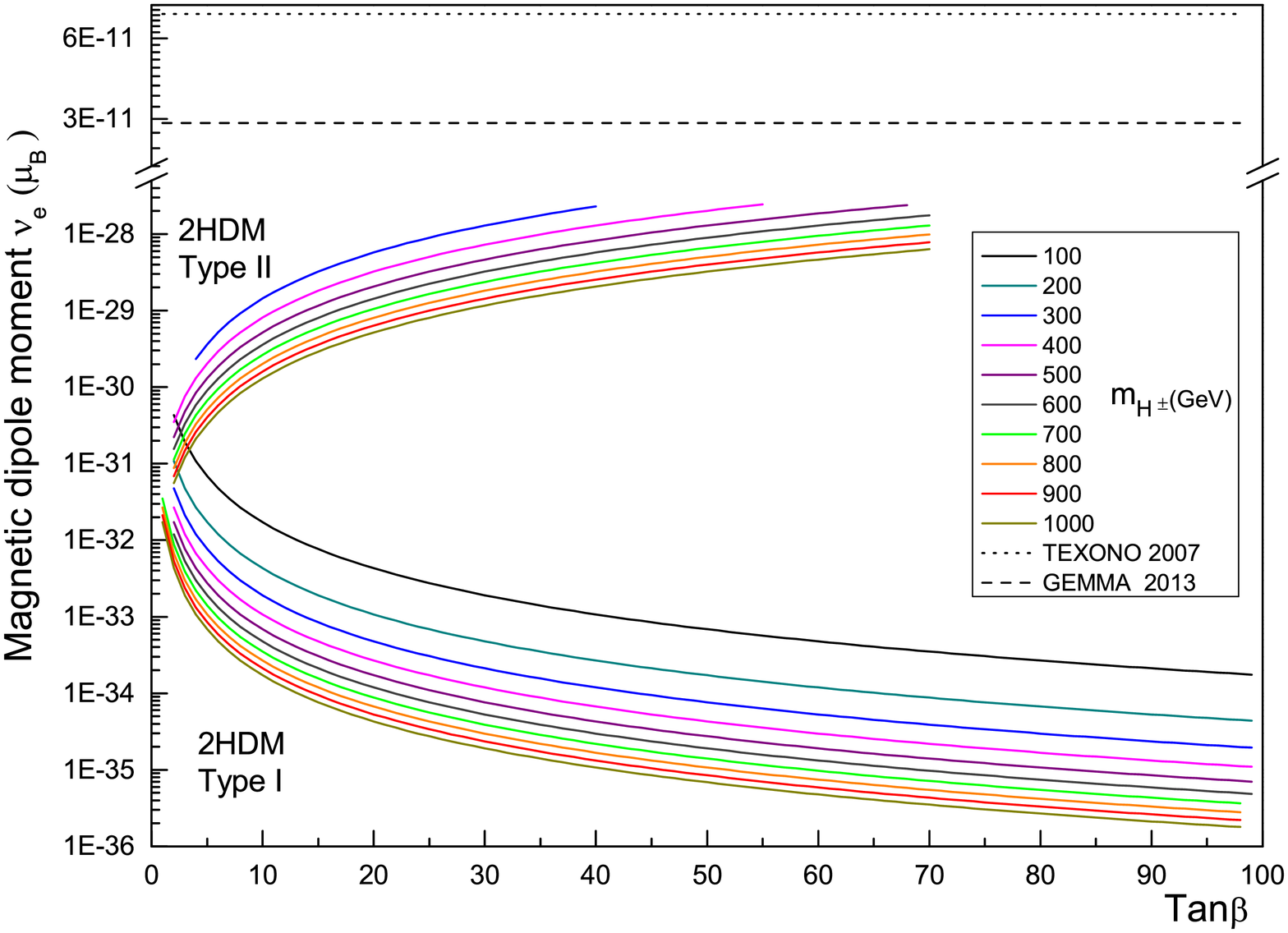} %
\includegraphics[width=0.48\columnwidth]{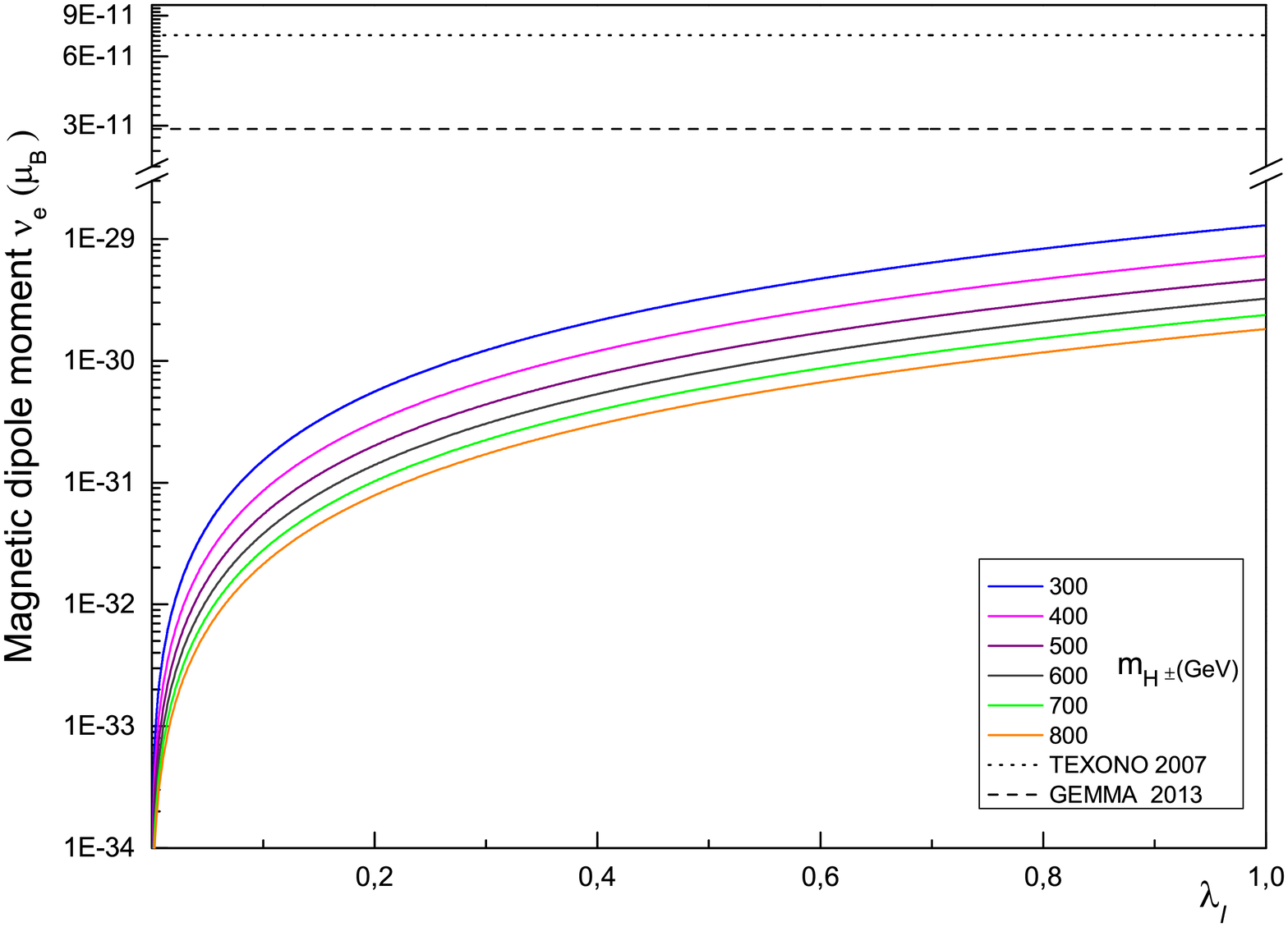} 
\vspace{0.3cm} \includegraphics[width=0.48\columnwidth]{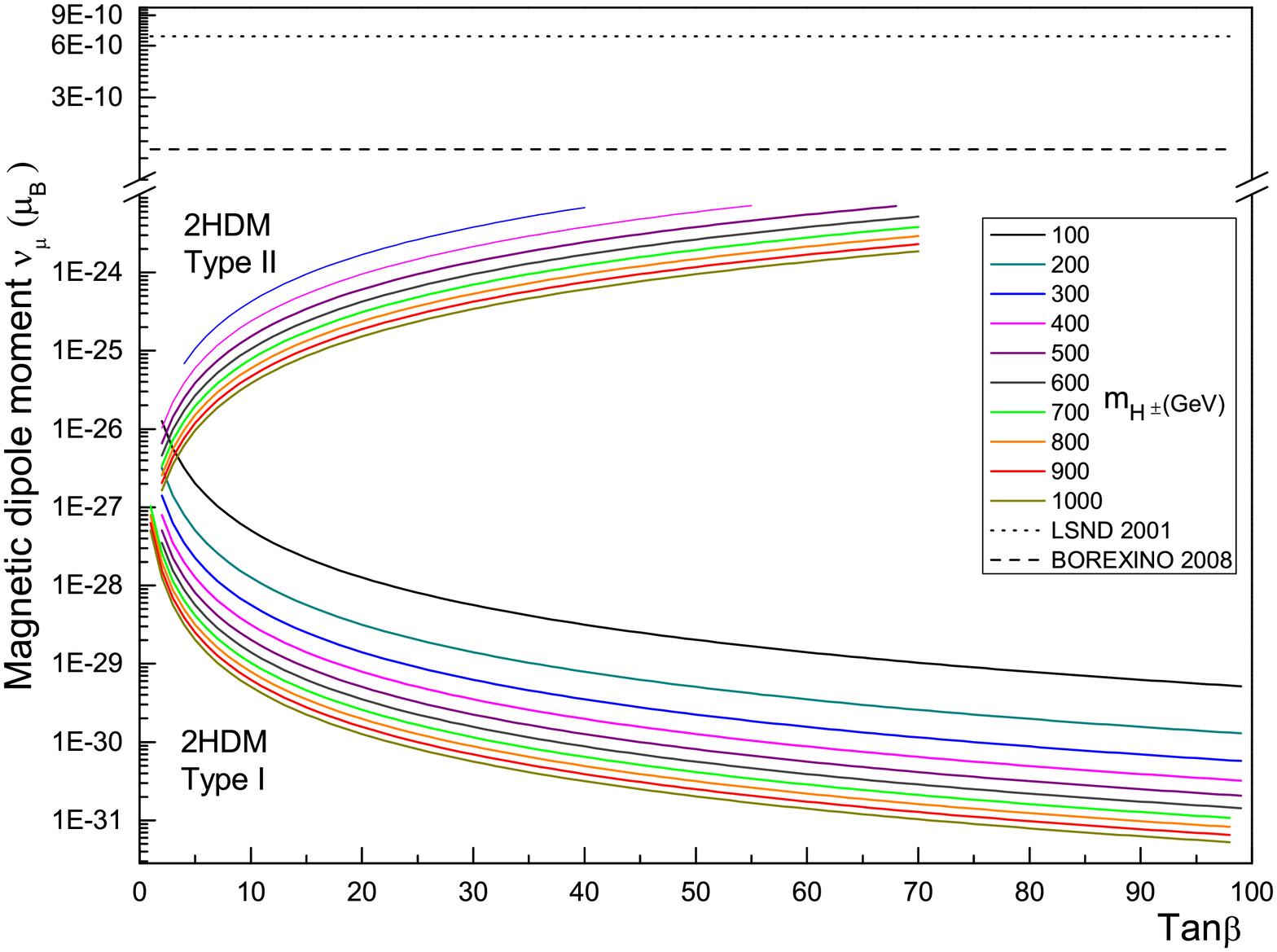} %
\includegraphics[width=0.48\columnwidth]{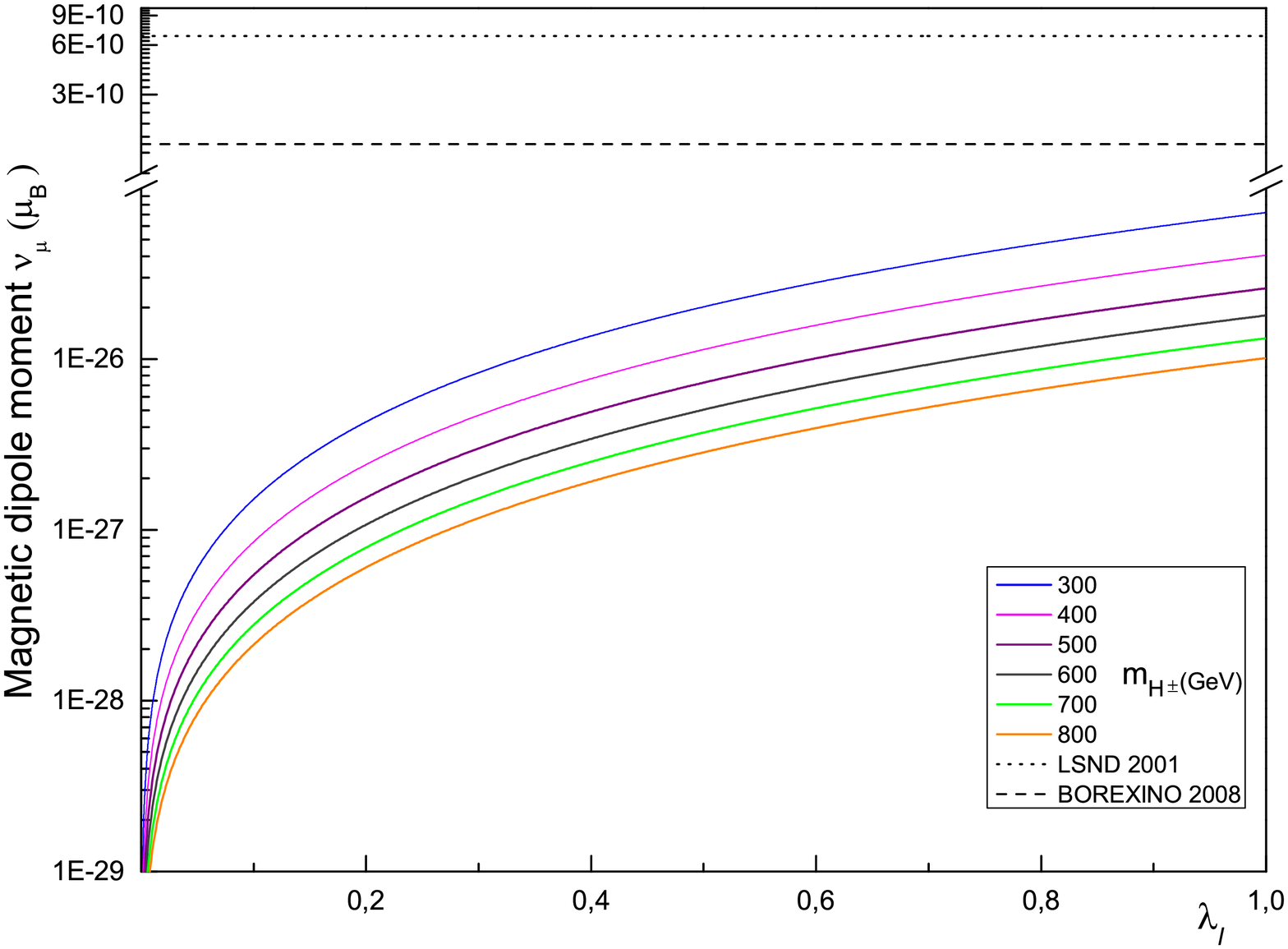} 
\vspace{0.3cm} \includegraphics[width=0.48\columnwidth]{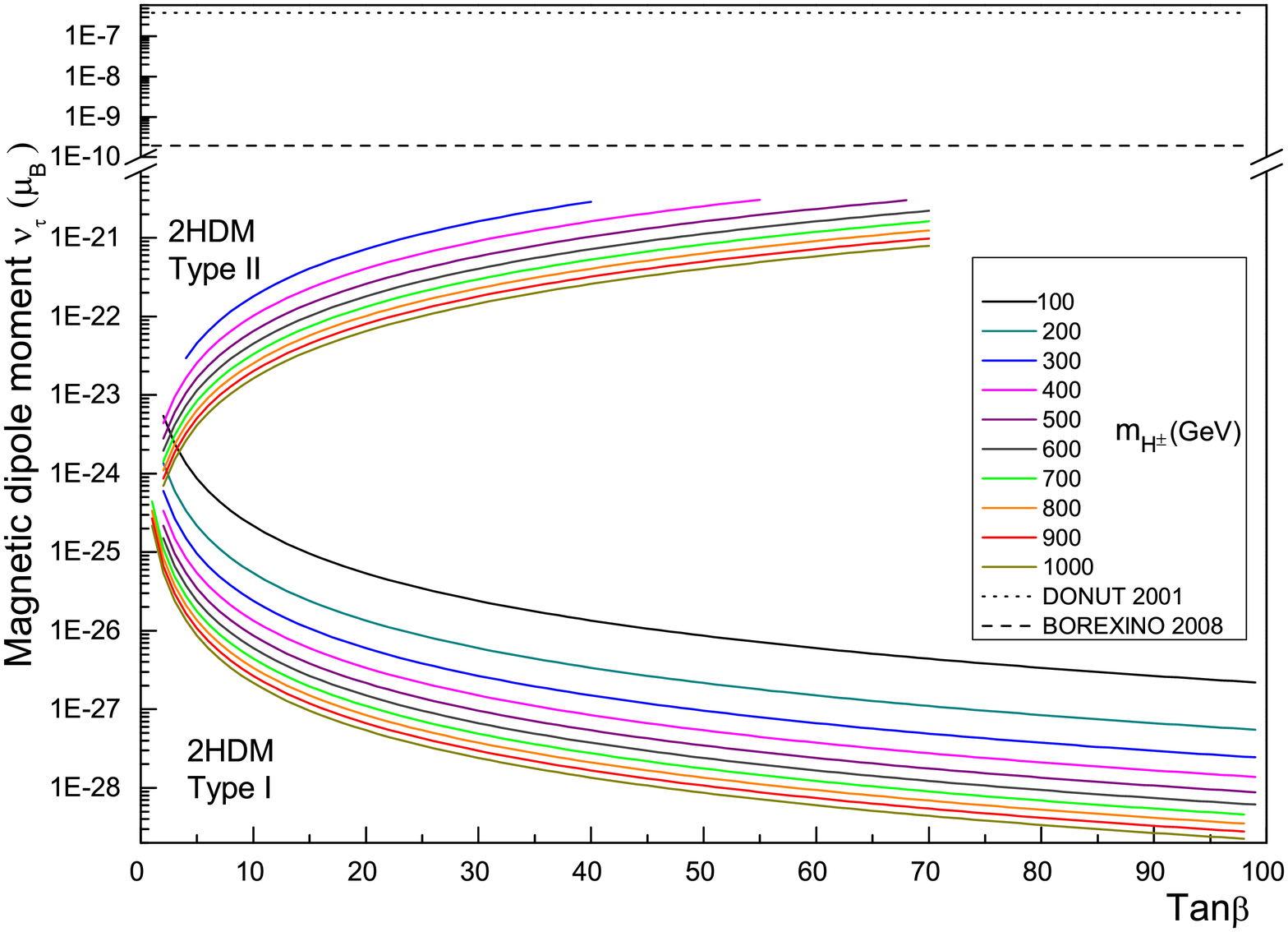} %
\includegraphics[width=0.48\columnwidth]{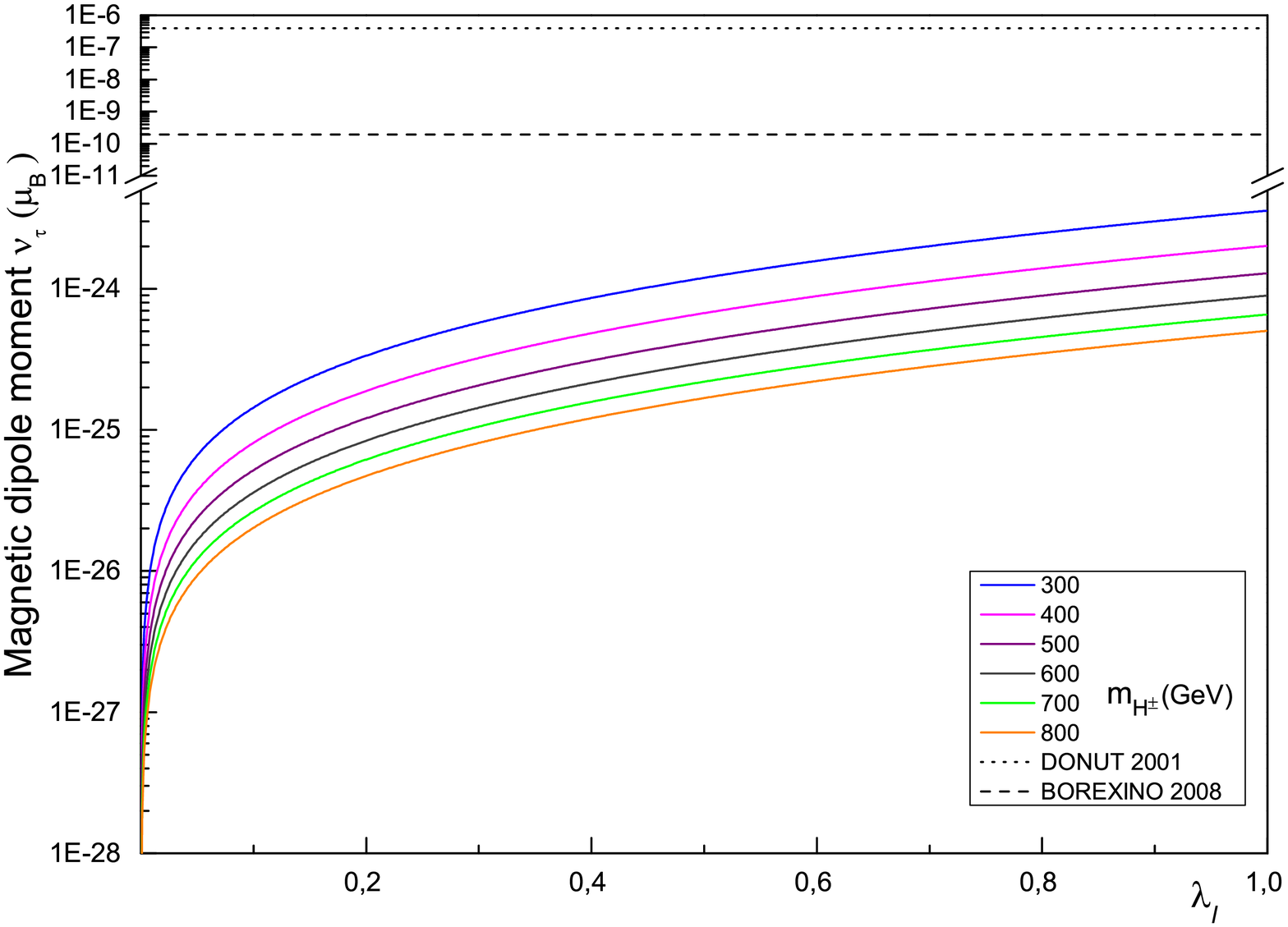} \vspace{%
-0.4cm}
\caption{\textit{(\textbf{Left}) Contribution to the magnetic dipolar moment in 
presence of magnetic fields for $\protect\nu_{e},\protect\nu_{\protect\mu},\protect\nu_{\protect\tau}$%
-neutrinos coming from type I (and Flipped), II ( and Lepton-specific) 2HDMs
with masses of charged Higgs sweeping between }$(100-900)$\textit{\ GeV to
type I, }$(300-900)$\textit{\ GeV to type II to different values of }$\tan 
\protect\beta $\textit{\ to each mass of charged Higgs. (\textbf{Right})
Contribution to the magnetic dipolar moment for neutrinos coming from type
III-2HDM. Here we have taken the masses of charged Higgs sweeping between }$%
(300-800)$ and $\protect\lambda _{\protect\nu },\protect\lambda _{l}\in %
\left[ 10^{-6},1\right] $. \textit{The horizontal dotted line makes
reference to the experimental thresholds for each neutrino flavor at $90\%$
C.L.. }}
\label{type_I_II}
\end{figure}

The contribution to the MDM due to the presence of magnetic fields is below
the SM contribution for all models (with and without natural flavor 
conservation). Furthermore, as our recent work has showed \cite{gomez}, 
contributions from magnetic fields are less than those obtained in the vacuum for 
the models considered. As it happens in analyses from vacuum for MDM, 
there is a strong relationship among MDM, neutrino and charged lepton masses. 
We can see this comparing the MDM contributions due to the tau neutrino regarding the muon and electron 
neutrinos in fig. \ref{type_I_II}.

\end{document}